\documentclass[conference]{IEEEtran}
\usepackage[utf8]{inputenc}
\usepackage[hyphens]{url}
\usepackage{cite}
\usepackage{amsmath,amssymb,amsfonts}
\usepackage{algorithmic}
\usepackage{graphicx}
\usepackage{textcomp}
\usepackage{csquotes}
\usepackage{rotating}
\IEEEoverridecommandlockouts

\def\BibTeX{{\rm B\kern-.05em{\sc i\kern-.025em b}\kern-.08em
    T\kern-.1667em\lower.7ex\hbox{E}\kern-.125emX}}
\begin{document}

\title{SoK: Uncentralisable Ledgers and their Impact on Voting Systems}

\author{\IEEEauthorblockN{Lionel Dricot}
\IEEEauthorblockA{\textit{Université catholique de Louvain}\\
\textit{ICTEAM/Crypto Group} \\
B-1348 Louvain-la-Neuve, Belgium \\
lionel.dricot@uclouvain.be}
\thanks{Lionel Dricot is partially supported by the European Commission and the Walloon Region through the FEDER project USERMedia (convention number 501907-379156)}
\and
\IEEEauthorblockN{Olivier Pereira}
\IEEEauthorblockA{\textit{Université catholique de Louvain}\\
\textit{ICTEAM/Crypto Group} \\
B-1348 Louvain-la-Neuve, Belgium \\
olivier.pereira@uclouvain.be}
\thanks{Olivier Pereira is partially supported by the SeVote project of the F.R.S.--FNRS}
}

\maketitle

\begin{abstract}
As we observe a trend towards the recentralisation of the Internet, this paper raises the question of guaranteeing an everlasting decentralisation. We introduce the properties of strong and soft uncentralisability in order to describe systems in which all authorities can be untrusted at any time without affecting the system. We link the soft uncentralisability to another property called perfect forkability. Using that knowledge, we introduce a new cryptographic primitive called uncentralisable ledger and study its properties. We use those properties to analyse what an uncentralisable ledger may offer to classic electronic voting systems and how it opens up the realm of possibilities for completely new voting mechanisms. We review a list of selected projects that implement voting systems using blockchain technology. We then conclude that the true revolutionary feature enabled by uncentralisable ledgers is a self-sovereign and distributed identity provider.

\end{abstract}


\section{Recentralisation of the Internet}
The Internet has been designed as a decentralised network. Each Internet server is a central authority node that has full control over the services it provides. Unlike a distributed network, where each node is independent, this kind of decentralised network still relies on multiple authorities\cite{Baran1964}. This topology is sometimes referred as "federated network" while "decentralised network" has evolved to cover both a federated or a distributed network\cite{Narayanan2012}.

It is important to note that technically federated or distributed networks are still considered as centralised if all nodes are managed by a single root of authority or trust\cite{Troncoso2017}.  We can infer that if some authorities in a distributed network control a large number of nodes, the network is mostly a federation of those authorities, even if technically looking distributed.

If the technical layers that allowed to build the Internet are decentralised, there is a natural social and economic incentive to recentralise, at least politically and economically\cite{Elias}. We observe that the Internet is currently moving toward centralisation\cite{Filippi2012}. 

It is now cheaper, more efficient and often technically more secure to run applications on Amazon S3 instead of your server, to post on Facebook instead of your website or to have your emails on Google instead of maintaining a mail hosting infrastructure. This recentralisation of the Internet in mega-silos can be witnessed as more and more traffic goes through Google and Facebook alone\cite{AndreStaltz2017}. The power of very few companies over the whole internet makes those companies as powerful as states, prompting the name "Net states"\cite{AlexisWichowski2017}. The control of the Internet by a few companies is not limited to user traffic. It goes as far as building their own Internet infrastructure\cite{KlintFinley2016}, including cross-oceanic cables\cite{DeborahBach2017a}.  For some, this recentralisation is even seen as the death of the Internet~\cite{FarhadManjoo2017}.

\section{Defining decentralised systems}

We use the term "political centralisation" to describe a system where the decisions are taken by one central authority, even if the system may be distributed from a technical point of view. Troncoso, Isaakidis, Danezis and Halpin \cite{Troncoso2017} suggest that technical distribution is not relevant in this context and that only political centralisation matters. They define a decentralised system as "\emph{A distributed system in which multiple authorities control different components and no single authority is fully trusted by all others}". In that context, an authority should be understood as a single entity taking decisions for a subset of the system thus effectively deciding how the components in this subset should behave. When the system is distributed, each user becomes its own authority, and the terms can be used exchangeably. In the rest of this paper, we will use "authority" to describe an independent entity which takes decision affecting a subset of the system. For a centralised system, the authority is the one in control. For a distributed system, authorities are the users. For a federated system, authorities are the nodes trusted by the users.

In a distributed network, users can choose who to trust, to which degree each node should be trusted or even decide not to trust anyone. In some cases, a decentralised system can also be used to reach an internal consensus without knowing which authorities to trust. This problem was called the Byzantine generals problem by Lamport, Shostak, and Pease\cite{Lamport1982}.  

In a federated network with few authorities, like today’s Internet,
you either have to trust some of the authorities in place or to
develop new mechanisms that allow running applications on an untrusted
infrastructure. Applied cryptography focuses on the latter choice,
allowing users to transmit information and compute through insecure
mediums, emulating the existence of trusted
party\cite{GMW87}. Cryptography enables an important paradigm shift in
our society: from trusting people or legal entities (because we have
no other choices) to trusting mathematics~\cite{Antonopoulos2014}.

Even when the network is technically decentralised, a federation of a small number of authorities opens the door to a potential recentralisation through collusions of those authorities. There is a strong economic incentive for collusion and recentralisation\cite{Narayanan2012}. But there are many cases where that centralisation is not desirable or is even dangerous\cite{Filippi2012}. Is the recentralisation unavoidable? Some decentralisation networks have shown a great resilience against centralisation, like the BitTorrent network which offers incentives to cooperation\cite{Andrade2005} despite several attempts to shut it down or at least control it to avoid its use to share copyrighted material.   

To further analyse the situation, we felt that the centralised/decentralised notion was not enough of a characterisation of the systems if we wanted to study the possibilities of online voting on a decentralised system. Indeed, we understood that a network could evolve and become centralised even though it was decentralised at some point. For online voting systems, we look for guarantees that hold in the long term. Everlasting privacy is an example of such requirement\cite{MN06}. We then looked for a characterisation of a system that would offer us a guarantee of everlasting decentralisation.

\section{Introducing uncentralisability}

Decentralised systems, as defined by Troncoso, Isaakidis, Danezis and Halpin\cite{Troncoso2017}, are only decentralised at one point in time. They are vulnerable to recentralisation through concentration of power. We say that a system is recentralised if one single authority has to be trusted by all users of the system. This means that this particular entity can take a unilateral decision that affects the whole system. Users are forced to accept the decisions as long as they want to use the system.
 
We foresee two ways to concentrate power in an initially decentralised system: by 1) collusion of authorities or by 2) disappearance of authorities. We formalise this notion by saying that a system $S$ is decentralised if it has $N$ authorities with $N \gg 1$. To ease the reasoning, we will use a simplification and consider that $S$ is decentralised if $N > 1$. In reality, we envision that the number of independent nodes needed to guarantee the decentralisation of a system will vary or even lead to a "degree of decentralisation" which is outside the scope of this paper.

We use the notation:
\begin{equation}
S(A_{1}, A_{2},\dots,A_{n})
\end{equation}
to indicate that a system $S$ is controlled by authorities $A_{1}, A_{2},\dots,A_{n}$.
We will often use the notation $A_{i\dots j}$ to refer to the set of authorities $\{A_i, \dots, A_j\}$. 

$S$ will become centralised if the authorities in $A_{1\dots n}$ collude or are controlled by a global authority $B$.  In which case, $N=1$ and $S$ is centralised.

\begin{equation}
\label{coal}
A_{1\dots n} \subset B \Rightarrow S(A_{1}, A_{2},\dots ,A_{n})=S(B)
\end{equation}

It is important to note that collusion of existing local authorities can be done in the open (acquisition, merge), be hidden (a single shadow authority control multiple local authorities) or happen through an oligopolistic agreement.

$S$ will also become centralised if $A_{2\dots n}$ disappear from the system. $N$ becomes $1$ and $S$ is centralised.

\begin{equation}
\label{disap}
S(A_{1\dots n} - A_{2\dots n})  = S(A_{1})
\end{equation}

A historical example includes the XMPP federated chat network. The network was fully decentralised, but Google Talk quickly became the most prominent actor. In 2013, Google decided unilaterally to leave the XMPP federation. Google users with google-only contacts saw no impact. But people using other servers than Google were suddenly forced to create Google accounts if they wanted to chat with their Google friends\cite{ParkerHiggins2013}. While the XMPP network survived and is still in use today, that single decision from a single authority had a deep impact on users, even those who didn't trust Google directly but had Google contacts. Users who trusted Google were suddenly in a centralised system. 

This example illustrates that decentralisation at one point in time gives no guarantee for the future. If decentralisation is critical, it should be guaranteed. For that reason, we introduce uncentralisability as a property of decentralisation guaranteed in time.

\emph{A system is uncentralisable if it can run and evolve without the participation or consent of any particular authority, even distributed.}

But is uncentralisability possible and provable? We will first explore a strict definition of uncentralisability through a notion that we call strong uncentralisability. We will then relax our definition to make it more practical and introduce soft uncentralisability.

\subsection{Strong uncentralisability }

We define a strong uncentralisable system as a system in which it is strictly impossible for a given entity to gain enough influence to control the system. To guarantee that impossibility, the system needs to be resilient against collusions of existing authorities and the disappearance of authority nodes. Strong uncentralisability can be seen as a perfect property, leading to an ideally decentralised system.

Resisting against collusion, as seen in (\ref{coal}) means that the more influence an authority has, the harder gaining influence should be. As collusion can be hidden, there should be an incentive for each individual authority to remain independent. A local authority should maximise its self-interest by remaining independent.

\begin{equation}
\label{anticol}
S(A_{1},A_{2}) \Rightarrow A_{1} \cap A_{2} = \emptyset 
\end{equation}

As some authorities may disappear, sometimes suddenly, as seen in \ref{disap}, another property of a strong uncentralisable system is that it should be easy to create a new authority that has immediate influence and could counter-balance other authorities.

\begin{equation}
\label{antidisap}
\exists A_{n+1} \nsubseteq A_{1\dots n} :  S(A_{1\dots n}) + S(A_{n+1}) = S(A_{1\dots n+1})
\end{equation}

The difficulty is that $A_{n+1}$ has to be created as soon as $N \leq 1$, which implies that the system must know $N$ at any time.

We observe that (\ref{antidisap}) is also a solution against collusion as seen is (\ref{coal}). If a strong uncentralisable system can exist, then all the authorities have to actively collaborate not to merge or to split/create a new authority as soon as the number of authority is too low. As the collusion might be external to the system itself, and thus undetectable by the system, decentralisation could be possible only if there is an incentive to do so. 

\emph{A system is said to be strongly uncentralisable if the incentive for each individual authority to remain decentralised is stronger than any incentive to political recentralisation. }

As the tendency to recentralise is mostly economic, it would make sense to build an uncentralisable system where there is an economic incentive to remain distributed. People working to keep the system distributed would be economically rewarded. But this idealisation fails to be resilient against Sybil attacks. A centralised authority may act like multiple decentralised authorities, taking centralised decisions without letting users know about it. Because of that, it might be impossible to guarantee the strong uncentralisability of a system or even to build such system. 

To make it more practical, we relax our requirements and introduce a soft uncentralisability notion.

\subsection{Soft uncentralisability}

Since a strongly uncentralisable system  could be overly demanding or even impossible to build, we realised that we could potentially infer some interesting properties of a system which, while not permanently decentralised, could be made decentralised at any time. We called that property soft uncentralisability. Unlike strong uncentralisability, which guarantees a permanent decentralisation, soft uncentralisability only guarantees decentralisation when needed.

\emph{A system is softly uncentralisable if it guarantees that any trusted authority can be untrusted at any time without affecting the system.} 

In a softly uncentralised system, the system may temporarily become centralised as long as every actor trusts the central authority. But it will always be possible to create a new authority without any interference from older authority. This means that the central authority keeps that position only as long as there is a complete trust from every actor. 

Interestingly, we realise that soft uncentralisability is equivalent to (\ref{antidisap}) but without the hard requirement of knowing $N$ at any time.   

Soft uncentralisability ($SU$) is thus a subset of Strong uncentralisability ($HU$). Strongly uncentralisable systems are softly uncentralisable, but the opposite is not true.

\begin{equation}
SU \subset HU
\end{equation}

A decentralised system might not be softly uncentralisable and, at some point, a softly uncentralisable system might not be decentralised. Soft uncentralisability is a notion orthogonal to decentralisation.

\subsection{Perfect Forkabilty}

Every decentralised system exists because there is a consensus agreed in the form of a protocol. If the consensus is broken, the decentralised system is forked in two new systems (which are either centralised or decentralised). Blockchains related projects usually differentiate backwards compatible fork ("soft fork") from non-backwards compatible fork ("hard fork")\cite{Bonneau2015}. As the XMPP example demonstrated, such a fork may impact the whole system\cite{ParkerHiggins2013}.

In the XMPP fork of 2013, one branch of the fork became centralised (the Google one) and not the other. Intuitively, this may be caused by the fact that each branch of the fork was very different. Users were already on one side of the fork before the fork happened. Would it be better if users could choose their branch after the fork? We define the notion of Perfect Forkability ($PF$). In a Perfectly Forkable system, each branch is treated in perfect equality. Each branch of the fork should continue to work exactly as before, with the only exception being the rule that created the disagreement leading to the fork. A user of a perfectly forkable system should be, after the fork, a user of both systems. 

\begin{equation}
\label{pf}
S(A_{1\dots n}) = S_{1}(A_{1\dots n}) \cup S_{2}(A_{1\dots n})
\end{equation}

If perfect forkability is guaranteed, it should nevertheless be rare enough and used only when no other solution is possible to avoid a fragmentation of the community. There should be an incentive to avoid forking, the convergence incentive. The balance here is subtle because convergence incentive and perfect forkability look contradictory.

This is not the case. A system might be perfectly forkable, but forks should only happen when there is a clear disagreement or a trust issue. Forks should not happen by accident but only when there is a clear decision to do it that counterbalances the convergence incentive.

We observe that if authorities have to choose one branch after a fork, (\ref{pf}) is equivalent to (\ref{antidisap}). This would mean that perfect forkability is a property of a system equivalent to soft uncentralisability. Intuitively, this makes sense: if $S(A_{1})$ became centralised and $A_{1}$ refuses to cooperate with a newcomer $A_{2}$, there will be a fork resulting in $S_{1}(A_{1}) + S_{2}(A_{1},A_{2})$. To be softly uncentralisable, a system has to be perfectly forkable. And if a system is perfectly forkable, it is softly uncentralisable.

An ideal strongly uncentralisable system has both soft uncentralisability and perfect forkability properties.

\begin{equation}
PF \equiv SU \subset HU
\end{equation}

Perfect forkability is usually impaired by the fact that one branch of the fork will keep the name and credibility of the initial project while the other branch has to build a new brand. In an ideal perfectly forkable system, there is no trademark allowing one authority to own the name of the project.

Open Source code in a git repository is an example of a  softly uncentralisable system.

Any user, even newcomer, could always contribute to the code. The ultimate authority might be centralised but only as long as everyone accepts it. In case of a disagreement, even a temporary one, the code will be forked, but each project may continue exactly as before (with the exception of the name, which is a reason that explains why Open Source projects trademark their name and logo). Every contributor of the initial project automatically becomes a contributor to both branches of the fork. Once some code has been released under an open source license, it becomes virtually impossible to take over all instance of that particular code.

In reality, the centralisation of the branding of a project does not appear to be a major problem which could prevent forking. It can be assumed that it is not impeding too much the soft uncentralisability property. As a way to circumvent that problem, forks use creative strategies: "Bitcoin Cash", a fork of Bitcoin, use a very similar logo and advertises itself as "peer-to-peer electronic cash"\cite{bitcoincash}, which is the title of the original Bitcoin whitepaper\cite{Nakamoto2008}. Bitcoin Cash users sometimes refer to it as "the true Bitcoin"\cite{CecilledeJesus2017}. "Ethereum Classic" is a name that speaks for itself for an Ethereum fork. It advertises itself as "a continuation of the original Ethereum blockchain"\cite{ethereumclassic}.

\section{Introducing the uncentralisable ledger primitive}

Now that we have defined uncentralisability, we consider that the Bitcoin blockchain is a clear attempt at building an uncentralisable ledger (if not one of the first). As pointed out by Narayanan et al.~\cite[p.31]{Narayanan2016}, with the current state of research, Bitcoin ironically works better in practice than in theory. It is thus important to understand and analyse the factors that come into play without relying blindly on the initial system proposal\cite{Bonneau2015}.

To simplify further analysis of the uncentralisability property in systems such as Bitcoin, we will introduce the "uncentralisable ledger" as a new ideal cryptographic primitive.

\emph{An uncentralisable ledger is a ledger which is softly uncentralisable. }

The direct consequence is that an uncentralisable ledger is perfectly forkable.

As the term "blockchain" has no strict definition and may be used on projects with varying levels of similarity with Bitcoin\cite{Narayanan2017}, introducing the "uncentralisable ledger" primitive gives us a framework to differentiate projects that relies on uncentralisability, such as Bitcoin and Ethereum, against other projects that may use a lot of properties from the blockchain technology but do not need the uncentralisability. This latter category includes many private or permissioned blockchains [21] where there is a clear central authority.

This new cryptographic primitive may also allow us to identify strong security weaknesses in projects that presuppose the uncentralisability property but use a ledger that has not been demonstrated to be uncentralisable.

\subsection{Properties of an uncentralisable ledger}

Bonneau, Miller, Clark, Narayanan, Kroll and Felten introduced a framework for characterising the stability of each layer of Bitcoin\cite{Bonneau2015} where stability means that the system will continue to behave in a way that facilitates a functional currency. As for the consensus protocol, the following properties have been identified to guarantee the stability of the system: Eventual consensus (the nodes will at some point agree), Exponential convergence (the probability of a fork of depth $n$ is $O(2^{-n})$), Liveness (new blocks may be added), Correctness (Only correct transactions are included in the longest chain) and Fairness (a miner with a proportion $\alpha$ of the total computational power will mine a proportion $\sim\alpha$ of blocks).

Those properties are specific to the Bitcoin blockchain. We will thus generalise those properties to an uncentralisable ledger:

\textbf{Immutability:} Any new entry in the uncentralisable ledger will eventually become impossible or exponentially hard to alter or to remove. This property is essential to build a ledger.

\textbf{Availability:} The entire history of the ledger might be accessed by any user at any time. If it is not the case, some user may restrict access to the system, creating a political centralisation. (It should be noted that some payload data might be encrypted and thus not readable by all the users. This is not in contradiction with the availability property as long as the encrypted data are not a structural part of the ledger). If this property is not met, a branch of a fork may have less information than the other branch, meaning that the ledger is not perfectly forkable and not uncentralisable.

\textbf{Correctness:} New entries to the ledger are guaranteed to respect the rules of the ledger. Incorrect entries are guaranteed to be found in a reasonable time. Rules of the ledger evolve through consensus of all the authorities.

\textbf{Perfect Forkability:} In case of a disagreement, the ledger might be forked. No branch of the fork are privileged, and users of the original ledger should automatically become users of both ledgers.

\textbf{Convergence:} The system tends to converge and has incentives to avoid accidental forks. If this property is not met, a ledger will easily split into multiple atomic ledgers, each of them becoming centralised. As said above, there is a subtle equilibrium between having perfect forkability and having incentives not to use it.

\section{Uncentralisability guaranteed by proof-of-work}
\subsection{Bitcoin economical incentive to uncentralisability}

As a technical concept, blockchains are mainly a distributed ledger mechanism. The concept has been studied and experimented before Bitcoin\cite{Narayanan2017}. Bitcoin introduced two new elements to the game.
Firstly, the bitcoin consensus protocol of which node should be added next was based on a proof-of-work algorithm known as mining. Miners were rewarded through the generation of bitcoins\cite{Narayanan2016}, p.105.
Secondly, bitcoins started being exchanged for fiat money. The exchange rate gave bitcoins an economic value. The reasons behind this economic value of Bitcoin can be discussed but was envisioned since the start as an essential incentive to guarantee the stability of the Bitcoin network\cite{Nakamoto2008}\cite{Bonneau2015}.
The original design of Bitcoin considers that the combination of those two elements is a strong economic incentive against centralisation\cite{Nakamoto2008}. Unlike other Internet layers, were centralisation is economically cheaper, on the Bitcoin blockchain, each actor has a strong economic incentive to keep the system decentralised.

Even before Bitcoin, Aspnes, Jackson and Krishnamurthy proposed a proof of work algorithm to mitigate Sybil attack on the Byzantine's generals problem\cite{Aspnes2005}. With Bitcoin algorithm, miners compete against each other to be the first to find the next block. Other than the raw mining incentive, people willing to publish a transaction in a block can also give incentives in the form of a transaction fee. Miner will then select the transactions with the highest fees and earn more.

The incentive for miners is thus clear: having the higher possible hash rate. And the higher the global hash rate is for the network, the harder it is to gain control over the blockchain and the harder the problem becomes. Increasing your bitcoin revenue through mining is exponentially harder: the more hash rate, the harder the problem becomes\cite{Narayanan2016}, p. 108.

The incentive that guarantees the uncentralisable property of the Bitcoin blockchain is economic. For each individual, it is economically suitable to not collude with others. We can thus conclude that Bitcoin is "human greed based technology". The technical infrastructure (Bitcoin’s blockchain) purposely exploits a generic trait of human psychology (greed) in order to guarantee a technical feature (uncentralisability). The rise of interest in blockchain technologies and the rise of the value of blockchain-based cryptocurrencies implies an unalterable trust toward human greed. 

\subsection{Is Bitcoin softly uncentralisable?}

Is Bitcoin strongly uncentralisable? Kiayias, Koutsoupias, Kyropoulou and Tselekounis demonstrated that a miner with more than 30 or 40\% of hashing power would be advantaged and find more than 40\% of the blocks if following a dishonest strategy\cite{Kiayias2016}. A trend towards centralisation has been observed in mining\cite{Beikverdi2015}, confirming that the proof of work algorithm used by Bitcoin is not strongly-uncentralisable. Centralisation in the Bitcoin mining business is well known and sometimes referred as ASIC-powered centralisation\cite{IanAllison2017}.

But what are the consequences of a single, possibly shadow authority having most of the hashing power? It means that that particular authority could be the only one able to propose new blocks to the blockchains. It should be highlighted that those blocks have to be valid. Thus, the central authority may block some transactions by not including them in any block or raise transaction fees\cite{bitcoinwiki}.

This is not seen as a major problem as long as the nodes of the Bitcoin network have the power to decide which code to run and which rules to follow\cite{Bonneau2015}. The particular implementation of the proof-of-work in Bitcoin might be compared to the legislative/executive separation of powers seen in most modern democracies. Nodes connected to the network decide which code to run (legislative power) while miners create new blocks that follow those rules (executive power).

It should be remarked that the number of nodes is not important: each node will agree with nodes that follow the same rules. If a node finds a longer blockchain with an incorrect block, it will dismiss it as incorrect\cite{bitcoinwiki}.

This makes controlling the nodes useless for any attack against the network itself. This is not true for attacks against individuals. Sybil attacks may be launched against a single node. If this node connects only to nodes controlled by an attacker, the attacker is in position to hide some blocks and thus launch a double-spending attack. This double-spending will eventually be found when the victim node eventually connects to an honest node. 

If the centralisation of mining becomes problematic, a proposal could be made to modify the proof of work algorithm, making current mining hardware useless. As soon as the nodes update to that particular algorithm, miners lose all their authority\cite{bitcoinwiki}. If there is a disagreement in the community, the system is forked. In theory, Bitcoin should be perfectly forkable. It was demonstrated to be the case in August 2017 with Bitcoin Cash\cite{bitcoincash}. Despite many concern about centralisation of mining\cite{B2014}, Bitcoin appears softly uncentralisable. The Bitcoin Wiki even considers the possibility of a manual fork of the code of the nodes to prune a malevolent branch of the blockchain, even though this chain would have required more computation power.

\begin{displayquote}[{\cite{bitcoinwiki}}]
"\emph{It's much more difficult to change historical blocks, and it becomes exponentially more difficult the further back you go. As above, changing historical blocks only allows you to exclude and change the ordering of transactions. If miners rewrite historical blocks too far back, then full nodes with pruning enabled will be unable to continue, and will shut down; the network situation would then probably need to be untangled manually (e.g. by updating the software to reject this chain even though it is longer).}"
\end{displayquote}

To follow the legislative analogy, we can say that the law is the protocol (or the source code). The nodes are the members of parliament who vote to modify the law. Miners are the executive power following the law. 

We conclude that Bitcoin soft uncentralisability seems guaranteed by a complex mix of separation of power and economic incentives, but a formal proof is yet to be found.

\section{Other consensus algorithms}

Proof-of-work is only one of the many consensus algorithms being currently developed. Proof-of-stake is one of the most popular alternatives which can take many forms, for example\cite{Kiayias2017}. The Algorand\cite{Gilad2017}, IOTA’s Tangle\cite{Popov2017} or proof-of-space\cite{Abusalah2017} are other consensus algorithms that aim to solve the Byzantine generals problem.

We believe that studying the uncentralisability of those algorithms and potentially proving formally their uncentralisability could help to understand them better and predict their resilience against potential recentralisation.

\section{Voting on a centralised infrastructure}

In a centralised institution, voting is a particularly sensitive matter. Most of the time, the central authority who organises the elections has a strong incentive to "win", even through cheating. In the same way that people want to transmit secure information through insecure mediums, voters want to ensure that the election is correct even though it is organised by an untrusted central authority. 

This has been the focus of the development of end-to-end verifiable systems, since the seminal works  of Chaum~\cite{Chaum1981}, and Benaloh and Fischer~\cite{CF85}. The extraordinary challenge addressed in these works is to build a system that protects voter privacy (only the voter knows her vote, apart from what could be derived from the election results) while enabling each voter to verify that his vote was cast as intended, recorded as cast, and that all and only votes submitted by authorised voters are properly counted.

While being fully verifiable, most of these systems include strongly centralised components, and election organisers could, alone or jointly, make an election fail. 

This is the case in traditional paper-based elections: an authority needs to provide a list of polling places and, if these polling places are later closed, then the election will simply not take place. This is clearly orthogonal to verifiability and, indeed, any observer would be able to verify that polling places are closed. A similar thing happens in most verifiable internet voting systems. For instance, in the Helios system~\cite{Adi08,ADPQ09}, an URL at which votes must be cast is provided as part of the election definition. If the authority maintaining this URL takes it down, then the election will fail. 

\section{Voting on an uncentralisable ledger}

At first glance, voting on an uncentralisable ledger instead of a centralised server would mainly help with transparency and auditability (counted as recorded)\cite{DanWallach2017}.  Another aspect improved by the use of an uncentralisable ledger would be robustness as it would become very hard for an attacker to take down the infrastructure.

Most Electronic Voting solutions require, at some point, a Bulletin Board in which public information can be added but never altered. Distributed ledgers, built specifically for a given election, where even considered as design for an electronic election system such as The Auditorium\cite{Sandler2007}.  

But none of those advantages makes an uncentralisable ledger a revolution in Internet voting. It could make it easier. But voting systems can be built and have been built without using this primitive.

We believe that the interesting question is: what are the possibilities enabled by an uncentralisable ledger that were impossible or considered as such before? Sociologists know that technology has a strong influence on the society we build. Building centralised technologies, such as nuclear power plant, lead to authoritarian and centralised states while decentralised technologies may lead to more decentralised societies with more individual freedom and autonomy\cite{Winner1980}. Wouldn't an incentralisable voting system deeply change the way we envision collective decision taking?

\section{New ways to apprehend an election}

The main properties of an uncentralisable ledger are immutability, availability, correctness, perfect forkability and convergence.

The availability and immutability properties allow reconsidering the timing of elections. Voting is not something done once in a while, with a campaign and planning, but can become permanent. After all, decisions have to be taken all the time in an ever-evolving context.

It is then possible to imagine a direct democracy system in which any voter has the right to write proposals and to submit them to the vote. Each proposal would have a deadline and would be marked as accepted by the community if it reaches a given amount of vote before the deadline. While such platforms already exist without an uncentralisable ledger, they always had to be run by a particular authority which had administrator rights on the platform.

This, of course, raises a lot of practical questions but the mere existence of an uncentralisable permanent and trustable infrastructure allows to think about new forms of governance or democracy\cite{Atzori2015}. In fact, those new form of government might even be urgently needed as there is an increasing conflict between the daily life of citizen (immediateness, self-sovereignty of actions) and the reality of a slow heavily centralised administration\cite{Mancini2015}. 

As an example, the Australian project VoteFlux\footnote{voteflux.org} proposed a liquid democracy system, including delegations, but with an incentive to not vote in order to transfer votes from unimportant issues to the more important one. Studying the implications of such voting system is outside of the scope of this paper, but it is interesting to note that their design (including voting token), while not explicitly referring to the blockchain, seems a lot easier to implement on a blockchain than on any other system. VoteFlux was implemented using Secure.Vote which is a blockchain based voting solution\cite{Lander2017}.

One of the features of some famous uncentralisable ledgers is the existence of tokens, which exist in a limited supply and which have an economic value.

Uncentralisable blockchains, valuable tokens and smart contracts allow the implementation of decentralised prediction markets. Such tools can be transformed into governance systems that implement a futarchist institution\cite{Hanson2013}. Another idea would be a voting platform linked to a decentralised reputation system, allowing people with more reputation to have more power, creating the first completely formalised meritocracy.

Another property which is interesting to consider is the perfect forkability. In case of a strong bipolar disagreement in a community, the community could fork. This, of course, raises the question of the physical infrastructure that cannot be forked but is an interesting lead for virtual communities. Being able to perfectly fork a community would probably change the way we envision conflict management and disagreement.

\section{Existing blockchain voting projects}

Nowadays, the term blockchain is widely used with various meanings. Multiple tech projects advertise themselves as "using blockchain" or being "blockchain related". The blockchain is also sometimes seen as the bread and butter, the silver bullet that will solve every problem and will make everything decentralisable. This belief sometimes also applies to e-voting where "voting on the blockchain" is seen as a miracle cure.

We have chosen to study existing public projects that advertise themselves as voting solutions (voting is the primary feature) based on the blockchain. We did not consider projects which may be implemented using blockchain-like technologies but did not advertise themselves as "blockchain based".

For each project, we investigated two aspects: 1) is the project built on an uncentralisable ledger and 2) is the project technically assuming uncentralisability.

We consider that by advertising themselves as "blockchain-base", those projects try to appear as "decentralised/uncentralisable" in the public eye. We can then analyse if 1) the project may be uncentralisable and 2) the project needs that uncentralisability.

It should be noted that new projects appear every day. This list should not be considered as exhaustive nor definitive.

\subsection{FollowMyVote}

FollowMyVote\footnote{https://followmyvote.com/} is an open source project developed by a US non-profit and implements different voting systems: Proportional Representation, Mixed Member and Majority.

The handling of elections is traditional and requires a central authority to handle identities and voter accreditation.

FollowMyVote implements its own blockchain. This has the unfortunate side-effect of making the blockchain easily centralisable. In the current state, FollowMyVote should be considered as a traditional centralised implementation using blockchain technology, not as an uncentralisable project.

\subsection{Procivis}

Procivis\footnote{http://procivis.ch} is commercial Swiss company that develops an e-voting solution in cooperation with the University of Zurich\cite{Procivis2017}. The Procivis solution uses the Ethereum blockchain which is softly uncentralisable.

While the counting is done on the blockchain, each voter should use a dedicated client to cast her vote. A prototype of one has been developed in Java but, in theory, voters could develop their own. Unfortunately, the smart contracts has to be deployed in a semi-private network, allowing a central authority to choose which nodes may run an instance of the voting software\cite{Matile2017}. This is an important point: being based on a uncentralisable ledger, in this case, the Ethereum blockchain, does not guarantee uncentralisability.

Code is announced to become open source. The Procivis solution relies on a central authority to provide ID and voter registration, in this case, the Swiss state. It should be noted that Procivis is also working on identity through eID+ and Valid\footnote{https://valid.global}, the later being blockchain based.

\subsection{Secure.vote}

Secure.vote\footnote{http://secure.vote/} implements a commercial liquid voting solution where voters can delegate their vote.

The system is blockchain agnostic and can be run on any blockchain. Initially built for VoteFlux, the project pivoted and is now currently focused on a specific niche: enabling a community to vote on the specific rules governing an Initial Coin Offering (ICO). With that specific use case in mind, Secure.vote makes a lot of sense when used on the blockchain of the project itself.

While the code is announced to be inspectable, it will not be free. Without unrestricted access to the source code, perfect forkability is impossible, guaranteeing that Secure.vote is centralised.

\subsection{The Open Vote Network}

The Open Vote Network\cite{McCorry2017} is a research project using smart contracts on the Ethereum platform. For performance reasons, the project is focused on small elections, such as boardroom elections. Identity of the voters must be known before hand.

It implements on the Ethereum blockchain the voting protocol called "Anonymous voting by two-round public discussion"\cite{Hao2010}. It should then be considered as a "classic" voting system which could be implemented without the blockchain. Code is Open Source\footnote{https://github.com/stonecoldpat/anonymousvoting}.

\subsection{Polys}

Polys\footnote{http://polys.me/} is a commercial voting project started by the private company Kaspersky. It uses a private Ethereum blockchain wich means that, while voting can be decentralised, there is no guarantee of uncentralisability.

While no code has been released yet, the project was announced to become Open Source.

\subsection{Votem}

Votem\footnote{https://www.votem.io/} is a commercial voting solution that claims to implement voting on a blockchain. From their documentation, it is clear that they use their own private blockchain, thus not using any uncentralisable feature.

It should be noted that no source code could be found nor any information about a potential release under an Open Source license. 

Votem announced an ICO. At the time of this writing, they did not release any whitepaper, ICO plan or source code.

\subsection{VOLT project}

The VOLT Project\footnote{http://www.volt-project.org/} is an academical research project that aims to explore the voting possibilities offered by the blockchain. No source code has been currently released. They first want to explore non-political voting scheme, such as shareholder voting. While such use cases are interesting, they do not take advantage of the uncentralisability of the blockchain.

\subsection{Democracy.earth}

Democracy.earth\footnote{http://democracy.earth/} is currently one of the most innovative projects when it comes to voting on a blockchain.

Based on the Ethereum blockchain, which is softly uncentralisable, the project implements a liquid democracy solution, where users can vote or delegate their vote. Users can also submit proposals. The identity of the voters is handled by the project itself in a decentralised way, making it a true standalone tool for flat governance.

Democracy.earth is the only studied project which seems to implement an uncentralisable voting solution, making it a compelling test case to study the impact of uncentralisability on voting and governance. The project is still under development and some concerns should be made, especially regarding the secrecy of the vote which is not perfectly guaranteed.

\subsection{Related Project: DCent}

DCent\footnote{https://dcentproject.eu} is an European funded multidisciplinary project exploring the future of democracy. The name itself is a reference to decentralisation. We did not include them on our short list because it uses the voting platform nVotes\footnote{https://nvotes.com}, previously known as AgoraVote, which does not use the blockchain and which is a centralised voting solution. 

It is interesting to note that DCent indirectly uses the blockchain through a crypto-asset called Freecoin\footnote{https://freecoin.dyne.org/} which is built to incentive participation in the democratic process. 

\subsection{Summary}

\begin{table}[htbp]
\caption{Summary of studied projects}
\begin{center}
\begin{tabular}{p{1.6cm} p{1.2cm} p{0.9cm} p{1cm} p{1cm}}
\begin{turn}{55}\textbf{Project name} \end{turn}&
 \begin{turn}{55}\textbf{\textit{Project Type}}\end{turn}& 
 \begin{turn}{55}\textbf{\textit{Open Source}}\end{turn}&
 \begin{turn}{55} \textbf{\textit{Uncentralisable Blockchain}}\end{turn}& 
 \begin{turn}{55} \textbf{\textit{Use Uncentralisability}}\end{turn} \\
\hline
FollowMyVote & Non-profit & Yes & No & No \\
Procivis & Commercial & Planned & No & No \\
Secure.vote & Commercial & No & No & No \\
Open Vote & Academic & Yes & Yes & No \\
Polys & Commercial & Planned & No & No \\
Votem & Commercial & No & No & No \\
VOLT & Academic & Unreleased & Not yet & Not yet \\
Democracy.earth & Community & Yes & Yes & Yes \\
\hline
\end{tabular}
\label{tab1}
\end{center}
\end{table}

Out of eight analysed projects claiming to be "blockchain based", only one ends up using uncentralisability as a core feature. Three of the projects even appear to be strongly centralised. Fortunately, no project assumed uncentralisability while using a centralised blockchain.

This result demonstrates the usefulness of the uncentralisability concept as it allows to differentiate projects and quickly identify important features.

There is nothing bad about using a centralised blockchain as long as the project itself is considered as clearly centralised. Projects based on an uncentralisable blockchain may, in the future, develop new features based on that uncentralisability. This is not the case for projects based on centralisable or centralised blockchains.

\section{Other governance solution related to the blockchain}

It is interesting to note that multiple projects address the governance problem without advertising themselves as voting platforms. They try to address the governance of collectivity from the point of view of project management solutions which is the kind of governance mostly seen in the industry. The concept is sometimes referred to as DAO (Decentralised Autonomous Organisation) and is perceived as a new form of organisation in direct competition with other institutions such as companies, markets, networks or even governments\cite{Davidson2016}. 

To illustrate this particular aspect, we selected three popular projects. This small list is not exhaustive and only serves as an example.

\subsection{Colony.io}

Colony.io\footnote{https://colony.io/} advertises itself as a platform for open organisations. The stated goal of Colony.io is to handle large projects with multiple stakeholders and multiple contributors. The platform includes project management features such a task management, a token used to reward users for their contribution and a reputation platform. 

Reputation is one of the core features of Colony.io. Reputation is earned in both "fields" and "skills" by completing tasks that are then judged by the task initiator.

Colony.io is permissive by default. Voting is seen only as a rare measure that should happen only when there is a dispute to settle. Initiating the dispute implies putting some token at stake, token that can be lost. The vote is meritocratic, the importance of a vote being pondered by the reputation of the voter. The voters share 10\% of the tokens at stake as a reward for taking part in the vote [17].

While the whole system seems highly dedicated to industrial collaboration, one example being the rLoop project\footnote{http://www.rloop.org/}, Colony.io exemplifies and codifies the meritocratic political process. 

\subsection{Backfeed}

Backfeed\footnote{http://backfeed.cc/} has similar foundations to Colony.io but with a more specialised approach toward reputation. In Backfeed, reputation is an essential tool that works as an incentive for voting.

Interestingly enough, reputation can be earned for voting early for the future consensus. Reputation is thus proportional to the ability to predict the outcome of a vote. While this is not explicitly mentioned in the Backfeed whitepaper (which is still being edited)\cite{Field}, this makes backfeed similar to some futarchist models.

Backfeed votes are public. There is not voter privacy.

\subsection{Aragon}

While it is philosophically similar to Colony.io and Backfeed, Aragon\footnote{https://aragon.one/} aims to create a global decentralised jurisdiction, hence encompassing the need for one single organisation. It could be seen as a network of organisations or an organisation of organisations.

The proposed system seems to be a mix between liquid democracy and futarchy, allowing token holders to make proposals and to vote, rewarding voters who voted for the accepted decision\cite{Cuende2017}. As with Backfeed, publicity of the votes is an essential tool to allow building reputation.

\section{The hard problem of identity}

An essential aspect of elections is the identity of the voter. Identity, possibly guaranteed through pseudonyms as proposed by Chaum~\cite{Chaum1981}, is needed to ensure that a given voter has the credential to vote and votes only once, as a way to guarantee political equality (even if, in some cases, these votes can carry different weights).

If the number of votes is not limited, the election becomes meaningless as the outcome is decided by the voters that have the most efficient voting strategy.

That’s why most electronic vote solution takes for granted the existence of a central authority issuing voting tokens to people allowed to vote. The existence of this token, which is implicit in the case of a list of registered voters, ensures that only allowed people vote and vote only once (or only the number of tokens they have received). A popular way of cheating the system is by using the token of a deceased person.

Of the explored voting solutions, only Democracy.earth addresses the identity problem. It is no coincidence if it is also the only project actively exploiting uncentralisability.

Several other blockchain related projects try to address the hard identity problem. They try to create a self-sovereign (each user owns and controls her own identity) and secure identity system based on an uncentralisable ledger. Evernym\footnote{https://www.evernym.com/}, Civic\footnote{https://www.civic.com/products/secure-identity-platform}, Uport\footnote{https://www.uport.me/}, Cryptid\footnote{http://cryptid.xyz/}, ShoCard\footnote{https://shocard.com/}, TrustStamp\footnote{https://truststamp.net/}, Democracy.earth or IDBox\footnote{http://www.idbox.io/} are examples of project trying to create an uncentralisable and self-sovereign identity solution based on the blockchain.

The World Food Program is currently experimenting with a blockchain-based identity platform for refugees. Called WFP Building Blocks, the project uses the Ethereum Blockchain\cite{WorldFoodProgramme2017}.

Analysing them is outside of the scope of this article. But the number of existing projects underlines a widely spread desire to create a decentralised identity solution and the strong belief that uncentralisable ledgers make it possible. It is interesting to note that some projects, such as WFP Building Blocks and IDBox, aim at solving the identity problem in situation when a central authority is not available, reliable or trustable.

The other strategy would be for pure voting systems like Procivis or Secure.vote to connect themselves to external identity providers. Such "Identity as a Service" providers probably make sense in the long term.

\section{The impact of governance of open source projects}

Even the most decentralised projects, like Democracy.earth, still have one big central authority: the source code. Special attention should be paid to the governance of the code itself. It seems intuitive enough that the code needs to be completely open source but even open source projects have various strategies of governance which have a series of trade-offs in terms of democratic governance, flexibility, and ability to evolve\cite{Wright2015}. One essential tool of open source governance is the fork, which allows any project to be split into two independents projects. When an uncentralisable blockchain related project is forked, this sometimes implies that the blockchain itself is forked, a phenomenon called "hard-fork".

There were hard-forks on the two main uncentralisable blockchains (Bitcoin and Ethereum), leading to the creation of entirely new assets (Bitcoin cash, Ethereum classic) with their own exchange value. In all the recorded cases, one branch of the fork continued to be considered as the canonical, initial project. But it is unclear what will happen if there is no consensus about what should be the canonical project after a hard fork, especially when the blockchain is used by applications such as voting and identity providing solutions.

Studying the impacts of coding governance strategies, including hard-fork, might be especially important for projects that are used to take collective decisions on a large scale.

\section{Conclusion}

In this article, we introduced several concepts: uncentralisability, soft uncentralisability, strong uncentralisability and perfect forkability. We reasoned that soft uncentralisability was equivalent to perfect forkability and was a property strong enough to guarantee the decentralisation of a system in the long term. One of the main findings is that a softly uncentralisable system might be centralised at some point in time.

We were then able to define a new cryptographic primitive called "uncentralisable ledger". We explored some properties of an uncentralisable ledger and argued that the Bitcoin's blockchain is an attempt at building such primitive.

While not strictly necessary, such uncentralisable blockchains may help to implement an electronic voting system. But they also enable the implementation of completely new voting paradigms.

We reviewed several existing electronic voting projects advertising themselves as voting solutions based on the blockchain. Using the uncentralisability concept proved to be useful to differentiate those projects and their potential for the future. From the reviewed projects, only Democracy.earth leverages the power of uncentralisable ledgers, making it the only project in the list that could not be implemented without an uncentralisable ledger.

Democracy.earth has the specificity of implementing its own self-sovereign and uncentralisable identity provider.  Identity is a key component of any voting system, and uncentralisable ledgers enable experimentations about building self-sovereign and uncentralisable identity providers, allowing complete self-governance without any central authority. 

As a last point, we also observed that any uncentralisable ledger software should be open source, open source projects being themselves softly-uncentralisable. We conclude by raising the point that studying the governance of open source projects behind an uncentralisable ledger is important.

\section*{Acknowledgment}

The authors would like to thanks, Mathieu Jamar for his help and his hindsight on the Bitcoin history and community.

\bibliography{library}{}
\bibliographystyle{IEEEtran}

\end{document}